\documentclass[preprint,12pt]{elsarticle}

\usepackage{amssymb}
\usepackage[dvips]{color}
\journal{JMMM}

\begin{document}

\begin{frontmatter}



\title{Finite size effects with variable range exchange coupling in thin-film Pd$/$Fe$/$Pd trilayers}


\author{R. K. Das, R. Misra, S. Tongay, R. Rairigh and A. F. Hebard}
\address{University of Florida, Department of Physics, Gainesville, Fl 32611}

\begin{abstract}
The magnetic properties of thin-film Pd$/$Fe$/$Pd trilayers in which an embedded $\sim$1.5$\AA$-thick ultrathin layer of Fe induces ferromagnetism in the surrounding Pd have been investigated. The thickness of the ferromagnetic trilayer is controlled by varying the thickness of the top Pd layer over a range from 8~\AA~to 56~\AA. As the thickness of the top Pd layer decreases, or equivalently as the embedded Fe layer moves closer to the top surface, the saturated magnetization normalized to area and the Curie temperature decrease whereas the coercivity increases. These thickness-dependent observations for proximity-polarized thin-film Pd are qualitatively consistent with finite size effects that are well known for regular thin-film ferromagnets. The critical exponent $\beta$ of the order parameter (magnetization) is found to approach the mean field value of 0.5 as the thickness of the top Pd layer increases. The functional forms for the thickness dependences, which are strongly modified by the nonuniform exchange interaction in the polarized Pd, provide important new insights to understanding nanomagnetism in two-dimensions. 
\end{abstract}

\begin{keyword}
Ferromagnetic palladium \sep Thickness dependent coercivity \sep Thickness dependent Curie temperature

\PACS 75.50.Bb \sep 75.50.Cc \sep 64.60.an


\end{keyword}

\end{frontmatter}


\section{Introduction}
The presence of 3$d$ magnetic transition metal ions in palladium (Pd) gives rise to giant moments thus significantly enhancing the net magnetization
\cite{ISI:A19611447C00007,ISI:A1960WG90100004,ISI:A19656314600010,ISI:A1973P242300045,ISI:A1975AP23700003,ISI:000175575100127,ISI:A19656404600017}. 
Pd is known to be on the verge of ferromagnetism because of its strong exchange enhancement with a Stoner enhancement factor of 
$\sim$~10\cite{ISI:A1986A628500027}. The magnetic impurities induce small moments on nearby Pd host atoms, thereby creating a cloud of polarization 
with an associated giant moment\cite{ISI:A1986A628500027,ISI:A19668279100017}. Neutron scattering experiments show that the cloud of induced moments 
can include $\sim$200 host atoms with a spatial extent in the range 10 to 50~\AA \cite{ISI:A19668279100017,ISI:000221426200053}. Thus a thin layer of 
Fe encapsulated within Pd will be sandwiched between two adjacent thin layers of ferromagnetic Pd with nonuniform magnetization and a total thickness 
in the range 20 to 100~\AA. 

We have investigated thin-film Pd/Fe/Pd trilayers in which the thickness $d_{Fe}$ of the Fe is held constant near 1.5\AA~ and the thickness of the 
polarized ferromagnetic Pd is varied by changing the top Pd layer thickness $x$. The magnetic properties are studied as a function of $x$. Our 
experiments are motivated by the recognition that ferromagnetism in restricted dimensions has attracted significant research interest and is not totally understood
\cite{ISI:000228837900194,ISI:000073242500003,ISI:000172938900018,ISI:000176350000014,Wolf1955}. For example, the coercive field $H_c$ increases 
as the thickness of the ferromagnetic film is decreased toward a thickness comparable to the width of a typical domain 
wall\cite{ISI:A1956WS77400018,ISI:A1958WB74500048}. The Curie temperature $T_c$ decreases as the thickness of the ferromagnetic film is 
decreased toward a thickness comparable to the spin-spin correlation length\cite{ISI:A1974T119000008,ISI:000255292500001,ISI:A1991GV08300036,ISI:A1992HE60300030,ISI:000167573800044,ISI:A1993LC51400083,ISI:A1994MW35100033}. Moreover, the
critical exponent $\beta$ increases with the thickness of the ferromagnetic film\cite{ISI:A1992HE60300030,ISI:A1994MW35100033}.
We will show below that similar phenomenology applies to ferromagnetically polarized Pd films, albeit with different functional dependences arising 
from the fact that exchange coupling, which decays with distance from the ferromagnetic impurity\cite{ISI:A1968C486500028}, is not uniform throughout 
the film.

\section{Experimental Details}

The samples were grown on glass substrate by RF magnetron sputtering. The base pressure of the growth chamber was on the order of 10$^{-9}$ Torr. First a thick layer of Pd of thickness 200~\AA~ is grown on top of the substrate. The root mean square surface roughness of this Pd layer was measured by atomic force microscopy to be $\sim$ 6~\AA. Then a very thin (1.5~\AA~ as recorded by a quartz crystal monitor) layer of Fe is deposited on top of the first Pd layer. Finally a top layer of Pd with thickness $x$ is grown to complete the trilayer structure shown schematically in Fig.~1a.

We discuss six different samples with the top Pd layer having a thickness $x$ varying from 8 to 56~\AA. The total thickness $y$ of the polarized Pd (see Fig.~1b) can range from 20 to 100 \AA\cite{ISI:A19668279100017,ISI:000221426200053}. Thus for $x < y/2$, changes in $x$ will give rise to changes in $y$. Auger electron spectroscopy (AES) was used to verify the presence of a well defined Fe layer. The AES measurements were performed in a $10^{-10}$ Torr vacuum at sequential intervals following removal of sub angstrom amounts of Pd using an Argon etch. The depth profile of the high intensity Fe3 (703.0~eV) LMM Auger electron peak of Fig.~1c shows that the Fe is embedded in the Pd as a distinct 2D layer with a FWHM thickness of 1.8~\AA. All of these steps were performed without breaking vacuum.

Measurements of the magnetization $M$ (Fig.~2) were performed using a Quantum Design MPMS system. The magnetic field $H$ was along the plane of the substrate. Since the magnetization measurements were \textit{ex situ}, $x$ was constrained to be greater than 8~\AA; otherwise the exposure of the sample to air caused unwanted oxidation of the Fe. The magnetic parameters $H_c(x)$ (Fig.~3) and $T_c(x)$ (Fig.~4) are calculated respectively from magnetization loops taken at 10~K (see inset of Fig.~3) and linear extrapolations of the temperature-dependent magnetization taken at $H = 20$~Oe (see inset of Fig.~4). The magnetic contribution from the bottom ferromagnetic Pd layer is independent of $x$, since $y/2 < 200 \AA$, the constant thickness of the bottom layer.
 
\section{Results and discussion}

For large values of $x$, the thickness $y$ of the combined polarized ferromagnetic Pd layers and the associated saturated magnetization $M = M_s$ will reach a constant value. This expectation is borne out in Fig.~2 which shows the $x$-dependence of saturated magnetization $M_{sA}$ normalized to sample area. We note that this normalized saturated magnetization $M_{sA}(x)$ increases with increasing $x$ as the total amount of polarized Pd increases. The onset of saturation, near $x = 30$~\AA~ indicates that the polarization cloud including the embedded Fe layer is $\sim$~60~\AA~ thick. This value is consistent with previous observation\cite{ISI:000221426200053}. The increase of $M_{sA}$ with $x$ shown in Fig.~2 is thus straightforward to understand. As $x$ increases the thickness of the top polarized ferromagnetic Pd layer increases with a concomitant increase of magnetic material in the system. Variation of $x$ clearly controls the thickness of the polarized ferromagnetic Pd layer. When normalized to the number of Fe atoms present, the saturated magnetization $M_{sA} = 1.1 \times 10^{-4}$emu/cm$^2$ corresponds to 9.2~$\mu_B$ per Fe atom, in close agreement with previous observations of the giant moment of Fe in Pd to be near 10 $\mu_B$\cite{ISI:A19668279100017}. 

Modeling the $x$ dependence of $M_{sA}(x)$ shown in Fig.~2 for our Pd$/$Fe$/$Pd trilayers is not straightforward. For regular ferromagnets with $M_s$ uniform throughout the thickness, we would expect $M_{sA}(x)$ to be linear in $x$; clearly it is not. A reasonable model will incorporate an exchange interaction $J$ that decays radially with the distance from the point ferromagnetic impurity\cite{ISI:A1968C486500028,ISI:A19656839900001}.
This complication requires modeling $J$ as a function of distance $x$ from the plane of impurity. A starting point would be to write the magnetization $M$ is a function of $J$\cite{AharoniBook1996}, 
\begin{equation} \label{FMequation}
 M(H,T,x)=M_sB_s \Bigg( \frac{M_s}{k_BT}\Bigg[g\mu_BH + 2pMJ(x) \Bigg] \Bigg)   ,
\end{equation}
where $B_s$ is the Brillouin function and $p$ is the number of the nearest neighbors beyond which $J$ is zero. In principle the experimentally determined values of $M(H,T,x)$ can be fit to Eq.~\ref{FMequation} to find the best fit values of $J(x)$ for different values of the parameter $p$. We have not performed such an analysis.


Fig.~3~ shows the behavior of the coercivity $H_c(x)$ as a function of $x$ (solid black circles). The data are well described by a power-law dependence (solid black line), $H_c(x) \propto x^{-\eta}$, where the exponent $\eta = 2.3(\pm 0.1)$ is close to the ratio 7/3. Similar power-law behavior reveals itself in regular ferromagnetic thin films where $\eta$ has a somewhat smaller value varying from 0.3 to 1.5\cite{ISI:000172938900018}. Because $\eta$ depends strongly on strain, roughness, impurity, and the nature of the domain wall (Bloch or Neel type)\cite{ISI:000172938900018}, it is not surprising to see a wide variation in $\eta$. Neel predicted for example that for Bloch domain walls, $H_c$ of a ferromagnetic thin film should vary as $x^{-4/3}$ when the thickness $x$ of the film is comparable to the domain wall thickness $w$ \cite{ISI:A1956WS77400018}. For the case of Neel walls, $H_c$ depends only on the roughness of the film and does not depend on film thickness \cite{ISI:000176350000014}. The variation of $H_c(x)$ becomes particularly pronounced when the film thickness becomes comparable to $w$. 

A qualitative understanding of the steeper $H_c(x)$ dependence becomes evident by recognizing that the formation of domain structure is driven by the reduction of long range magnetostatic energy which at equilibrium is balanced by shorter range exchange and anisotropy energy costs associated with the spin orientations within a Bloch or Neel domain wall. Domain wall thickness is given by $w = \sqrt{A/K}$\cite{ISI:000255292500001,ISI:000183656300003} where $K$ is the crystalline anisotropy constant and $A$ is the exchange stiffness, proportional to the exchange energy, $J$\cite{Turek2001}. The domain wall size $w$ increases for decreasing $K$ and increasing $J$. If $K$, which depends on the relatively constant spin-orbit interaction\cite{AharoniBook1996} within the Pd component of the Pd$/$Fe$/$Pd trilayers, remains constant, then variations in $w$ are dominated by variations in $J$. Thus as $x$ decreases toward zero, the increase in $J$\cite{ISI:A1968C486500028} gives rise to an increase in $w$ which in turn gives rise to a more rapid increase in $H_c$ than would be seen in regular ferromagnets with constant $J$. As discussed above, this rapid variation with $\eta \sim 7/3$ is observed experimentally. 

The data in Fig.~4 show that $T_c$ increases as $x$ increases and reaches a relatively constant value near $x = 20$~\AA. The dashed black line is a 
guide to the eye and is qualitatively similar to the behavior of $M_{sA}(x)$ shown in Fig.~2 which saturates at a larger value near 30~\AA. These 
observations are again qualitatively consistent with the finite size effect associated with critical phenomena in 
ferromagnets\cite{ISI:A1974T119000008,ISI:000255292500001,ISI:A1991GV08300036}. Although the data are not of sufficient quality to distinguish the 
power-law behavior that is predicted for finite size effects\cite{ISI:A1974T119000008,ISI:000255292500001,ISI:A1991GV08300036}, we expect that the 
dependence is further complicated by the previously discussed dependence of $J$ on $x$ in polarized ferromagnetic Pd. The behavior of $T_c(x)$ 
suggests that Pd$/$Fe$/$Pd trilayer should be treated as a single layer with a well defined spin-spin correlation length. If the Pd layers are treated 
separately, then the bottom layer with fixed thickness $y/2$ would have a $T_c$ equal to the highest $T_c$ of the top layer. In this case the overall 
measurement would not show a strong change in $T_c$ as a function of $x$, since the $T_c$ of the bottom layer would dominate for all $x$. 

We note that for our planar geometry, $T_c$ decreases with decreasing thickness as has also been shown for thin-film Ni \cite{ISI:A1974T119000008} and 
epitaxial thin-film structures based on Ni, Co and Fe\cite{ISI:000255292500001}. On the other hand $T_c$ increases with decreasing size of 
ferrimagnetic MnFe$_2$O$_4$ nanoscale particles with diameters in the range 5-26~nm\cite{ISI:A1991GV08300036}. This increase of $T_c$ with decreasing 
size is attributed to finite size scaling in three dimensions where all three dimensions simultaneously collapse\cite{ISI:A1991GV08300036}. In our 
two-dimensional planar thin films only one of the dimensions, the thickness, collapses and $T_c$ decreases rather than increases in accord with the 
observations of previous studies\cite{ISI:A1974T119000008,ISI:000255292500001}. 
  
The order parameter (magnetization) is found to be linear when plotted as a function of $(1-T/T_c)^{\beta}$ as shown in Fig. 5 for all 6 samples. The data are taken in a small applied magnetic field of 20 Oe. 
The linear behavior is observed over a broad range of temperatures below $T_c$. The exponent $\beta$ is determined from the slope of the straight line when magnetization is plotted as a function of $(1-T/T_c)$ on logarithmic scales. As shown in the inset, $\beta$ increases smoothly with increasing $x$. This behavior again is qualitatively in agreement with regular ferromagnetic thin films\cite{ISI:A1992HE60300030,ISI:A1994MW35100033} where $\beta$ also increases with thickness. In the case of polarized ferromagnetic Pd we find $\beta$ $\sim$ 0.5. This is larger compared to regular ferromagnets\cite{ISI:A1992HE60300030,ISI:A1994MW35100033}. The value of $\beta$ varies from 0.44 for small $x$ and approaches  the mean field value of 0.5 as $x$ increases
(inset of Fig. 5). The mean field value is generally not seen in experiments for regular ferromagnets, instead $\beta$ is found to be either $\sim$ 0.125 
for 2D spin systems or  $\sim$ 0.325 for 3D spin systems\cite{ISI:A1992HE60300030,ISI:A1994MW35100033}. Our observation of mean field behavior is not understood, since the Pd/Fe/Pd system is complicated due to the non-uniform exchange coupling, $J$, which depends on the distance from the Fe layer. 

\section{Conclusions}
In conclusion, we have characterized the magnetic properties of thin-film Pd$/$Fe$/$Pd trilayers and determined that critical size effects apply 
to ``ferromagnetic'' Pd where the ferromagnetism is induced by proximity to an underlying ultrathin Fe film. The critical size, or equivalently 
the critical thickness, is controlled by varying the thickness $x$ of the top Pd layer. The dependences on film thickness of the coercive field 
$H_c$, the Curie temperature $T_c$ and critical exponent $\beta$ are in qualitative agreement with finite size effects seen in regular ferromagnetic films where the exchange 
coupling $J$ is constant throughout the film. The results presented here increase our understanding of nanomagnetism in ultrathin systems by showing  
that the spatial variations of $J$ in proximity coupled Pd have a pronounced influence on the form of thickness-induced variations, 
namely: a nonlinear dependence of $M_{sA}(x)$, an unusually strong power-law dependence of $H_c(x)$, a rapidly saturating dependence of $T_c(x)$ and a mean field 
critical behavior of magnetization ($\beta$ $\sim$ 0.5).

This research was supported by the NSF under grant number DMR-0704240 and as a part of the NSF Nanoscale Interdisciplinary Research Team project under grant number DMR-0403480.
We thank E. Lambers , D. Kumar and P. Kumar for useful discussions and technical assistance.
\label{}


\newpage
\bibliography{FePd}

\newpage
\begin{figure}
 \centering
 \includegraphics[width=15cm,angle=0]{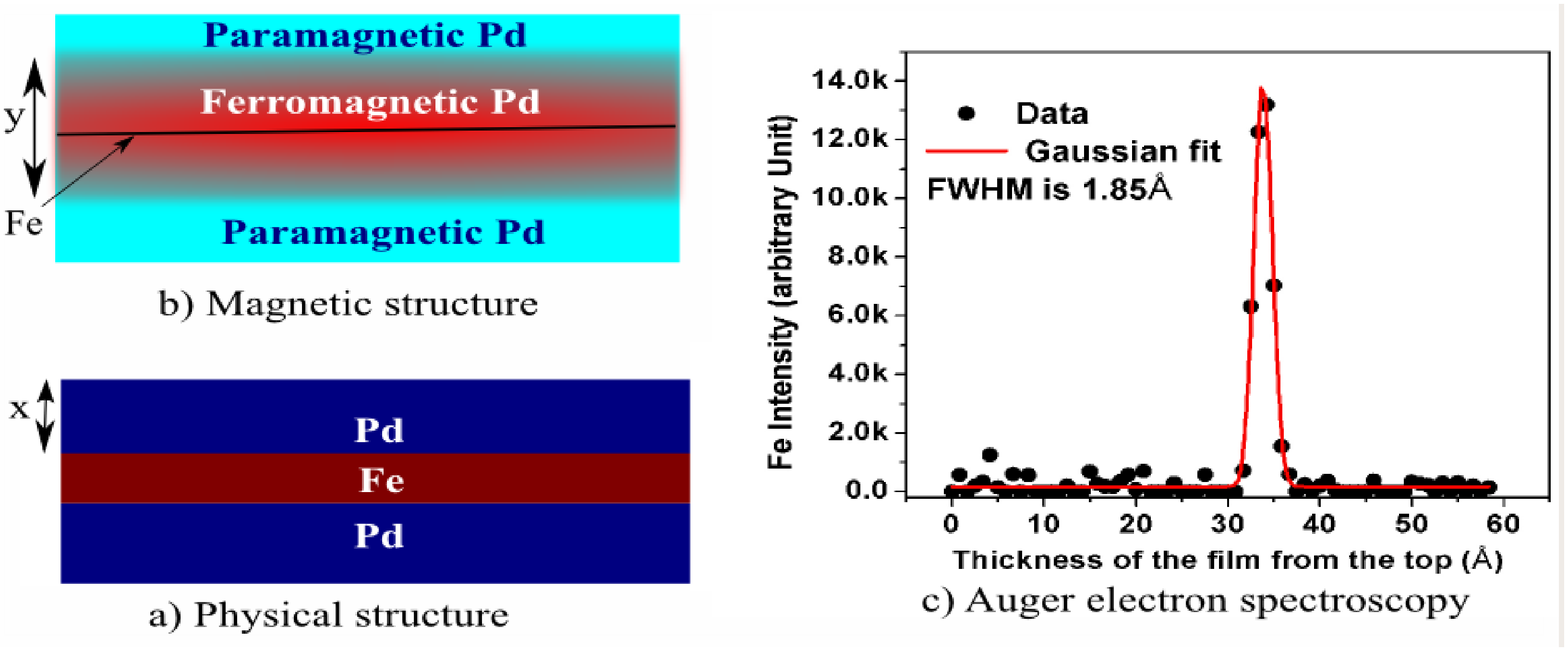}
 \caption[fig:sample]{a) Multilayer structure of a Pd/Fe/Pd trilayer. The bottom layer of Pd is 200~\AA~ thick. The thickness of the Fe layer is 1.5~\AA~ as recorded by the quartz crystal monitor. The thickness $x$ of the top layer of Pd is varied from 8 to 56~\AA. b) Magnetic structure of the sample. The total thickness $y$ of polarized Pd is in the range 20 to 100~\AA~(shaded red area). Thus by varying $x$, it is possible to vary the thickness $y$ of the polarized ferromagnetic Pd layer. c) Intensity of Fe3 (703.0~eV) LMM Auger electron peak plotted as a function of material removed by argon sputtering. The data (solid black circles) are fit to a Gaussian distribution (red line). The full width half maximum value of 1.85~\AA~is consistent with crystal monitor measurements.}
\end{figure}

\newpage
\begin{figure}
 \centering
 \includegraphics[width=12cm,angle=0]{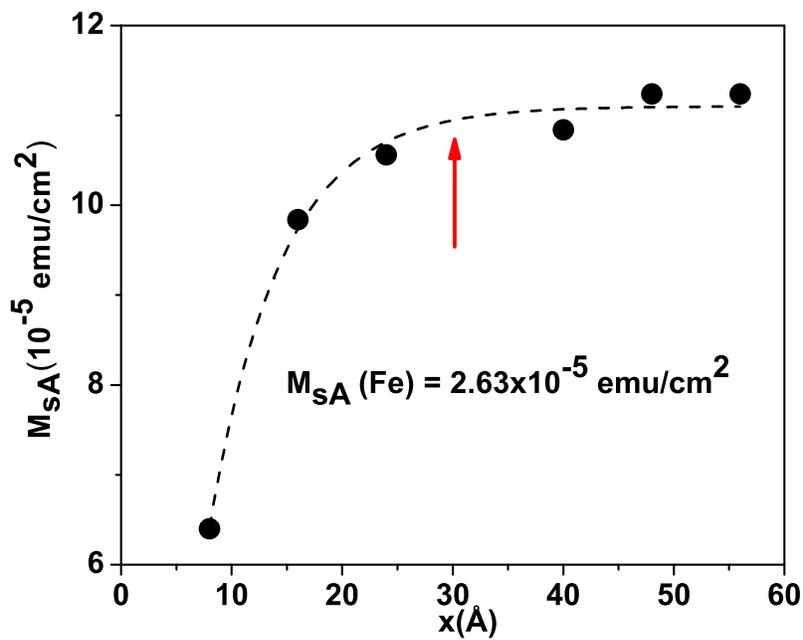}
 \caption[fig:msx]{The saturation magnetization normalized to the area of the sample $M_{sA}$ shows a smooth increase with increasing thickness $x$. The experimental data are shown as solid black circles and the dashed black line is a guide to the eye. Saturation to a constant value occurs near 30\AA~(vertical arrow).} 
\end{figure}

\newpage
\begin{figure}
 \centering
 \includegraphics[width=12cm,angle=0]{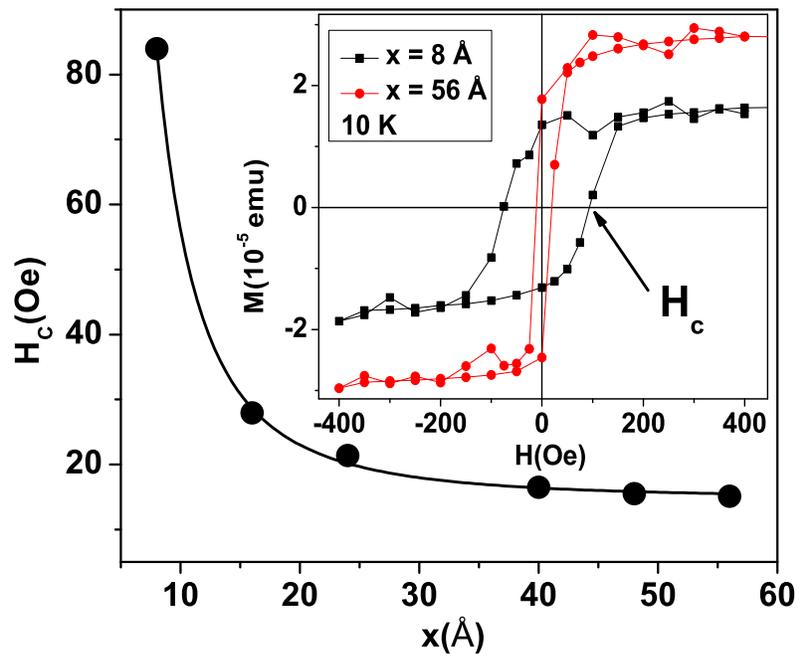}
 \caption[fig:hcx]{The coercive field $H_c$ shows a strong increase as the thickness $x$ of the top layer of the Pd decreases. The data are shown as solid black circles and the black solid line is a power law fit with exponent $\eta = 2.3(\pm 0.1)$. The inset shows magnetization loops at $T = 10$~K for $x = 8$\AA~ (solid black squares) and $x = 56$\AA~ (solid red circles).}
\end{figure}

\newpage
\begin{figure}
 \centering
 \includegraphics[width=12cm,angle=0]{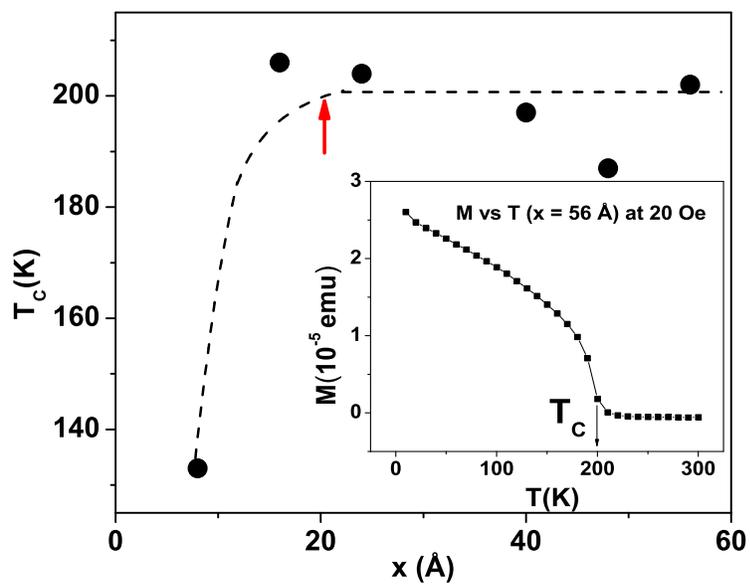}
 \caption[fig:tcx]{The Curie temperature $T_c$ rapidly increases with increasing $x$. Data are shown as solid black circles and the dashed black line is a guide to the eye. Saturation to a constant value occurs near 20\AA~(vertical arrow) The inset with $T_c$ indicated by the vertical arrow shows the temperature-dependent magnetization taken in a field $H = 20$~Oe.}
\end{figure}

\newpage
\begin{figure}
 \centering
 \includegraphics[width=12cm,angle=0]{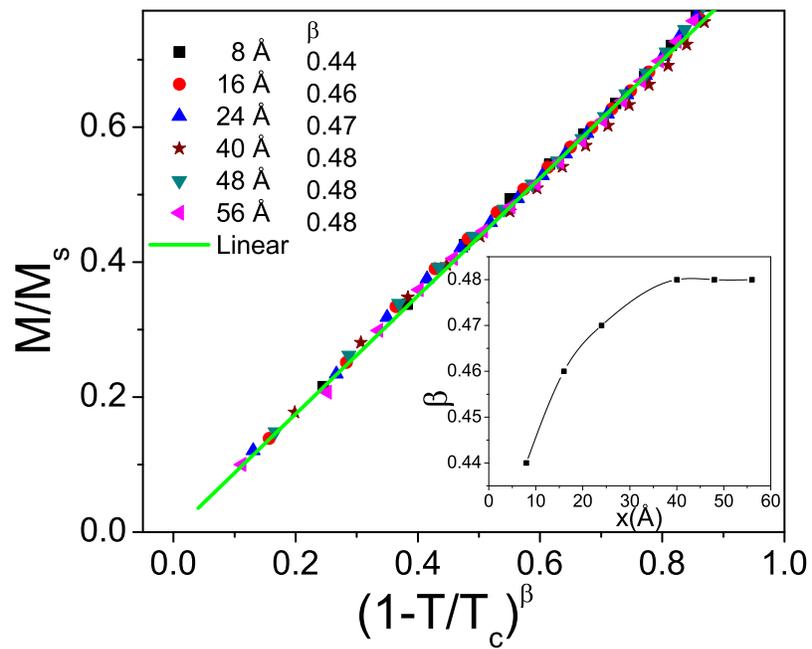}
 \caption[fig:beta]{Behavior of magnetization as a function of $(1-T/T_c)^{\beta}$ where $\beta$ is listed in the legend for each film. The inset shows that $\beta$ increases slowly and approaches the mean field value of 0.5 for large $x$.}
\end{figure}

\end{document}